\documentstyle[preprint,aps]{revtex}
\begin{document}
\draft
\title{QED corrections to Planck's radiation law and photon thermodynamics\\}
\author{M. Hossein Partovi\cite{email}}
\address{Department of Physics and Astronomy, California State University\\
Sacramento, California 95819-6041}
\date{\today}
\maketitle
\begin{abstract}
Leading corrections to Planck's formula and photon thermodynamics arising
from the pair-mediated photon-photon interaction are calculated.  This
interaction is attractive and causes an increase in occupation number for
all modes.  Possible consequences, including the
role of the cosmic photon gas in structure formation, are considered.
[hep-th/9308089]
\end{abstract}
\narrowtext

\section{INTRODUCTION}

Electromagnetic  radiation  in thermal equilibrium is a fundamental physical
system  which has played a crucial, if serendipitous, role in opening up new
frontiers  in physics; few systems are as ubiquitous or have played
as significant a role in the
progress of modern physics and astronomy.  Indeed it
was in an attempt to describe the properties of cavity radiation that Planck
stumbled  upon  the  notion of quantization in 1900, and, 65 years later, it
was  while  testing a microwave antenna designed for communications research
that   Penzias   and  Wilson  discovered  the  cosmic  microwave  background
radiation,  the  single  most  important  clue  to  the  past history of the
universe.   The  recent  discovery by Smoot and coworkers [1] of large-angle
temperature  fluctuations  in  the background radiation, which, by contrast,
had long been anticipated, promises to be another milestone in the quest for
the  origins  of  the  universe.   The distribution law discovered by Planck
accurately  describes  the  equilibrium properties of an assembly of photons
over  a  vast  range  of  temperatures  and  scales, from terrestrial cavity
radiation  (from  which  it was deduced) to hot stellar atmospheres, and, of
course,  including  the  2.73K  cosmic  background radiation.  Nevertheless,
there  are  small,  high-temperature  corrections to Planck's law and photon
thermodynamics which, in view of the pervasive presence of radiation systems
in general and the singularly important role of the background radiation for
cosmic  evolution  in  particular,  may play a significant role in an as yet
undiscovered  phenomenon.   For  this  reason,  as  well  as their intrinsic
interest  as  high temperature modifications to a fundamental law of nature,
these  corrections  are  worthy of serious attention.  This communication is
devoted  to the calculation of these corrections and the exploration of some
of their consequences.

The  Planck  law  basically  describes  a  noninteracting  gas  of massless,
spin-one  bosons (not subject to number conservation), driven to equilibrium
by  interaction with an external system, e.g., the atoms in the walls of the
cavity in the case of cavity radiation, and the charged particles present in
the  cosmic  fluid  in  the  case  of the
background radiation.  The virtual
nonexistence  of an interaction between photons under ordinary conditions, a
hallmark  of  Maxwell electrodynamics and the working principle of virtually
all  existing  telecommunication  systems, is what sets the photon gas apart
from  other  ideal gases.  To highlight this difference, let us consider the
fact that the mean free path for hydrogen under standard conditions is about
$10^{-6}$   m,   allowing  hydrogen  to  behave like a continuous medium on
scales larger
than  a  micron.   By  contrast,  the  same  quantity  for a photon gas is a
staggering  $10^{61}$  m  [cf. Eq. (19) below], implying that under ordinary
conditions  photons  behave like a collection of independent, free-streaming
particles on any physically meaningful scale.

The   absence   of   interaction   between  photons  and  the  linearity  of
electrodynamics are of course violated by the process of pair creation, a
fact  that was realized by Halpern [2] as early as in 1933.  Not long before
this, Delbr\"uck [3] had suggested that quantum effects would cause
photons to be scattered by
an external electric field.  Delbr\"uck's phenomenon, which was observed
in the scattering of 1.33 Mev gamma rays by the Coulomb field of the lead
nucleus some years later [4], is essentially the interaction of a real
photon  with  a  virtual  one [5].  By contrast, the interaction of two real
photons predicted by Halpern has never been
detected because of its extreme weakness.  The center-of-mass photon-photon
cross section is equal to $.031\alpha^{2}r_{e}^{2}(\omega/m)^{6}$ for
$\omega/m\ll1$ and nearly equal to it almost up to the pair production
threshold $\omega=m$ [6]; here $\alpha$ is the fine-structure constant,
$r_{e}$ the classical electron radius, $\omega$ the photon energy, and $m$
the  electron  mass  (natural  units  with $\hbar=c=k=1$ will be used unless
otherwise stated).
For 633 nm light, this cross section is about $4\times
10^{-64}$ cm$^2$, and it rises steeply to about $3\times
10^{-30}$ cm$^2$ near the production threshold.  These minute cross sections
are  the reason why phenomena involving the interaction of real photons have
not
been much discussed in connection with nonlinear electrodynamic effects,
attention having been focused
instead on the magnetic counterpart of Delbr\"uck scattering and other
effects  [7].   It  is  reasonable  to  expect,  however,  that
 these  tiny
magnitudes  will  eventually  succumb  to  high-precision experiments, or to
new observations.  Indeed a scheme has already been proposed for detecting the
photon-photon interaction by {\it colliding laser beam interferometry},
using a modification of Michelson interferometers developed for
gravity wave detection [8].

The present discussion deals with photon
interactions in equilibrium distributions, the  opposite  extreme to
highly-ordered,  far-from-equilibrium distributions characteristic of laser
light.
While at low (i.e., $\ll m$) temperatures a photon gas is
basically pure, at sufficiently
high  temperatures  a significant contamination of
electron-positron  (and  possibly  other)  pairs may  be  present, making
it inappropriate to
treat  the  gas as a single-component system.  To ensure that the photon gas
is  effectively  free  of  contamination, we shall restrict our treatment to
temperatures  for  which  the  photon-photon  collision  rate  dominates the
collision  rate of the photons with the pair-produced electrons or positrons
so that the latter play no significant role in the thermodynamics of the
photon  gas.   We  shall  see  below  that  the condition $T \leq 1.5 \times
10^{8}$ K, where $T$ is the temperature of the photon gas, is sufficient for
this purpose.  Now for this range of temperatures, the parameter $T/m$ is no
larger  than  $2.5  \times  10^{-2}$ and can be treated as small.  Since, for
dimensional  reasons, the   dependence of the corrections we are
seeking on $T$ must  be  through  the combination $T/m$, we can achieve
a considerable simplification in the
calculations by limiting them to the leading order in this parameter, with
essentially no loss in accuracy.
Our objective, therefore, is to
derive the thermodynamics of a photon gas (i.e., the charge-zero, equilibrium
state  of  QED) up to and including the leading correction in the parameters
$\alpha$ and $T/m$.

The leading contribution to photon-photon scattering is of second
order in
$\alpha$ and originates in the box diagram in which four external photon
lines
are attached to a closed electron loop [9].  In leading order, the matrix
element for
this diagram is directly related to the interaction energy of a pair of
photons, so that the corrections in question will be of relative order
$\alpha^{2}(T/m)^{4}$, the power of 4 resulting from the four electron lines
of  the  box  diagram  in  the  limit $T/m \rightarrow 0$.  Obviously, these
corrections will only be manifested under extreme conditions, requiring very
high  temperatures or very large spatial dimensions.  Such conditions may be
achievable  in  controlled  experiments,  or  may  have prevailed during the
evolution  of the universe subsequent to the annihilation of positrons.  For
example,  the  elastic  properties of the cosmic photon gas may have played
a role in the early stages of density perturbation growth and structure
formation .  As
will  be  shown  below, the cosmic background radiation has in the past been
under  conditions  such  that  the photon-photon collision rate exceeded the
expansion   rate.   This  fact  implies  the  possibility  of  independently
propagating  sound waves or supporting growing density perturbations for the
photon  gas.   Accordingly,  the  results derived below will be explored in
some  detail  with a view to their cosmological implications, although there
do not appear to be any observable consequences at this time.

\section{PHOTON-PHOTON INTERACTION ENERGY}

The  thermodynamics  of  a  photon  gas  can  be conveniently derived in the
canonical  ensemble  where the density matrix and the partition function are
given by
\begin{equation}
{\hat{\rho}}(\beta ,V) = Z^{-1}(\beta ,V) \exp [-\beta {\hat{H}}(V)],
\eqnum{1}
\end{equation}
\begin{equation}
Z(\beta ,V) = tr\{\exp [-\beta {\hat{H}}(V)]\}.
\eqnum{2}
\end{equation}
Here $V$ is the volume and $\beta = T^{-1}$.  To the accuracy sought here,
the Hamiltonian may be written as
${\hat{H}}={\hat{H}}_{0}+{\hat{H}}_{\text int}$, where ${\hat{H}}_{0}$
describes noninteracting photons and
${\hat{H}}_{\text int}$ is the effective interaction Hamiltonian
resulting from the box diagram.  Recall that our objective
is to calculate ${\hat{\rho}}$ in the limit $T/m \rightarrow 0$.
Since the amplitude resulting from the box
diagram is convergent, this limit can be implemented by
taking the static limit, $m \rightarrow \infty$, of ${\hat{H}}_{\text int}$
in Eqs. (1) and (2), a step that considerably
simplifies the calculation.  One simply calculates the amplitude for
the box diagram in the static limit, then converts this into
${\hat{H}}_{\text int}$, in much the same way that the Born amplitude is
converted into an interaction potential when dealing with potential
scattering.

A convenient alternative to the above procedure is to take advantage of
a well known result, known as the Heisenberg-Euler Lagrangian [10], that
recasts
the static limit of the box diagram amplitude as an
effective interaction Lagrangian expressed in terms of
the field variables.  The density corresponding to this Lagrangian is
\begin{equation}
{\hat{\cal L}}_{\text int}({\hat{\bf E}},{\hat{\bf B}})
= {2 \alpha^{2}\over
45m^{4}}[({\hat{\bf E}}^{2} - {\hat{\bf B}}^{2})^{2} +
7({\hat{\bf E}} \cdot {\hat{\bf B}})^{2}].
\eqnum{3}
\end{equation}
A simple calculation shows that with ${\hat{\cal L}}_{\text int}$ treated
to first order, the interaction Hamiltonian density is given by
${\hat{\cal H}}_{\text int} = -{\hat{\cal L}}_{\text int}$, i.e.,
${\hat{\cal H}}$ is equal to ${1\over 2}({\hat{\bf \Pi}}^{2}
+ {\hat{\bf B}}^{2})
-{\hat{\cal L}}_{\text int}({\hat{\bf \Pi}},{\hat{\bf B}})$.
Note that the canonical momentum density ${\hat {\bf \Pi}}$ is now equal to
${\hat{\bf E}}+\partial {\hat{\cal L}}_{\text int}({\hat{\bf E}},
{\hat{\bf B}}) / \partial {\hat{\bf E}}$,  the  second  term  being  the
effect  of the
interaction [11].

The next step is to calculate the diagonal matrix elements of the
interaction Hamiltonian
${\hat H}_{\text int}$ in the photon number representation (these being the
only ones needed in the following calculations).  It is convenient to carry
out this calculation in the Coulomb gauge using the complex field ${\hat
{\bf G}}={\hat {\bf \Pi}}+i{\hat {\bf B}}$.  The plane-wave expansion of
${\hat {\bf G}}$ in finite-volume normalization appears as
\begin{eqnarray*}
{\hat {\bf G}}(t,{\bf x})=&&{\sum}_{{\bf k},\lambda}{i k \over \sqrt
{2kV}} {\bf f}_{\lambda}({\hat {\bf k}}) \\
&&\times \left[ {\hat a}_{{\bf k},\lambda}
(t) \exp (i{\bf k}\cdot {\bf x})-{\hat a}_{{\bf k},\lambda}^{\dag}
(t) \exp (-i{\bf k}\cdot {\bf x}) \right],
\end{eqnarray*}
where $k=\vert {\bf k} \vert$ and ${\hat {\bf k}}= {\bf k}/k$.  Here
the unit vectors ${\bf e}_{\lambda}({\hat {\bf k}})$, $\lambda=1,2$,
represent the two polarization directions normal to $\bf k$, and
the complex vector ${\bf f}_{\lambda}({\hat {\bf k}})$ is given by
${\bf e}_{\lambda}({\hat {\bf k}})
+i{\hat{\bf k}}{\bf \times} {\bf e}_{\lambda}({\hat {\bf k}})$.  Furthermore,
${\hat a}_{{\bf k},\lambda}$ and
${\hat a}_{{\bf k},\lambda}^{\dag}$ are standard destruction and creation
operators subject to the commutation relations
\[
\left[ {\hat a}_{{\bf k},\lambda}(t),{\hat a}_{{\bf k}',
\lambda'}^{\dag}(t) \right] ={\delta}_{{\bf k},{\bf k}'} {\delta}
_{\lambda,\lambda'}.
\]
It is useful to note here that the product ${\bf f}_{\lambda}({\hat
{\bf k}}) \cdot {\bf f}_{\lambda'}({\hat {\bf k}}')$ vanishes for
${\hat {\bf k}}={\hat {\bf k}}'$.

We now proceed to calculate the diagonal matrix elements of $({\hat{\bf G}}
^{2})^2$ and ${\hat{\bf G}}^{2}
({\hat{\bf G}}^{\dag})^{2}$.  Let $\vert n \rangle$ represent the state with
$n({\bf k},\lambda)$ photons
of momentum
${\bf k}$ and polarization ${\bf e}_{\lambda}(\hat{\bf k})$.
Then, following a lengthy but essentially straightforward
calculation, we find
\[
\langle n \vert ({\hat{\bf G}})^{2} ({\hat{\bf G}}^{\dag})^{2} \vert n \rangle
= {2 \over V^{2}} {\sum}_{1,2} k_{1}k_{2} \vert {\bf f}_{1} \cdot {\bf f}_{2}
\vert^{2} n_{1}n_{2},
\]
where the sum runs over all possible values of $({\bf k}_{1},{\lambda}_{1})$
and $({\bf k}_{2},{\lambda}_{2})$, and the abbreviations
${\bf f}_{i}= {\bf f}_{{\lambda}_{i}}
({\hat{\bf k}}_{i})$ and $n_{i}= n({\bf k}_{i}, {\lambda _{i}})$ have been
introduced.  The corresponding matrix
element for $({\hat{\bf G}}^{2})^2$ is obtained from this formula by
replacing $\vert {\bf f}_{1} \cdot {\bf f}_{2}
\vert^{2}$ with $({\bf f}_{1} \cdot {\bf f}_{2})^{2}$.  Note that
in these sums all terms with ${\hat{\bf k}}_{1}= {\hat{\bf k}}_{2}$ vanish as
a result of the vanishing property of ${\bf f}_{1} \cdot {\bf f}_{2}$
mentioned above.

The photon-photon interaction energy can now be written down using the above
information; the result is
\begin{equation}
\langle n \vert {\hat H}_{\text {int}} \vert n \rangle =-
{\alpha^{2}\over
45Vm^{4}}{\sum}_{1,2}n_{1}, n_{2} k_{1}k_{2}
(4{\cal R}^{2}+7{\cal I}^{2}),\eqnum{4}
\end{equation}
where ${\cal R}$ and ${\cal I}$ are the real and imaginary parts of
the
quantity ${\bf f}_{1}{\cdot}{\bf f}_{2}$.  Note that
$\langle n \vert {\hat H}_{\text {int}} \vert n \rangle$
represents a sum of negative
two-body interaction terms; i.e., {\it photons attract photons}.  Furthermore,
both ${\cal R}$ and ${\cal I}$ vanish for
${\hat{\bf k}}_{1} = {\hat{\bf k}}_{2}$ as a result of the vanishing of
${\bf f}_{1} \cdot {\bf f}_{2}$ mentioned above; i.e.,
{\it parallel photons don't interact}.  The latter property can be
traced to the fact that the center-of-mass energy of a pair of parallel photons
vanishes, implying the same for the corresponding cross section.

\section{PHOTON THERMODYNAMICS}

To  first  order  in  the  interaction, the density matrix of Eq. (1) can be
represented as
\begin{equation}
{\hat{\rho}} = Z^{-1}\int_0^1 d \xi \exp (-\xi \beta {\hat H}_{0} )
(1-\beta {\hat H}_{\text {int}} ) \exp[(\xi-1)\beta {\hat H}_{0} ],\eqnum{5}
\end{equation}
from which the partition function can be calculated using the result
given in Eq. (4);
\begin{equation}
Z = {\sum}_{\{n\}}(1-\beta \langle n \vert {\hat H}_{\text {int}}
\vert n \rangle)
\exp(-\beta \langle n \vert {\hat H}_{0} \vert n \rangle),\eqnum{6}
\end{equation}
where $\langle n \vert {\hat H}_{0} \vert n \rangle$ is the energy of an
assembly of noninteracting photons, given by the familiar
expression ${\sum} n({\bf k},{\lambda})k$.  The symbol
$\{n\}$ in Eq. (6) indicates an unrestricted summation (since photon number is
not fixed) over $n({\bf k},{\lambda})$ for each possible value of
$({\bf k},{\lambda})$.

Thermodynamic  quantities  can  now be calculated as ensemble averages using
Eqs. (5) and (6).  For example, the mean energy $U(\beta,V)$, defined by
$tr({\hat H}{\hat \rho})$, can be calculated by applying
the formula $-(\partial / \partial \beta) \ln Z$.
To find $U$, we first substitute
(4) in (6) and perform the sum over $\{n\}$ to get
\begin{equation}
Z = \left[1-\beta  {\sum}_{1,2}\ g(1,2) {\bar n}_{0}(k_{1})
{\bar n}_{0}(k_{2})\right] Z_{0},\eqnum{7}
\end{equation}
where  $g(1,2)$  is  the  quantity  multiplying the photon numbers in Eq. (4),
${\bar n}_{0}(k)$ is  the  mean  occupation  number
for  noninteracting  photons as given by
the Planck  formula $[\exp(\beta k)-1]^{-1}$, and $Z_{0}$  is the associated
partition  function  ${\prod} [1-\exp(-\beta k)]^{-2}$.   The product in the
last expression runs over all possible values of ${\bf k}$.

To complete the calculation of $U$, we find it convenient to replace $g(1,2)$
in (7) by its average, ${\bar g}(k_{1},k_{2})$, over the two polarization
directions
as well as the orientations of
the  momenta (a permissible operation since ${\bar n}_{0}$
only depends on $k$); the result is
$-22 \alpha^{2} k_{1} k_{2}/135 V m^{4}$.
Using  this  result  in  (7) and carrying out the remaining sums (in
the large-volume, continuum limit), we find
\begin{equation}
Z = \left[1 +{22 \alpha^{2} \beta V u_{0}^{2} \over 135 m^{4}}\right]Z_{0},
\eqnum{8}
\end{equation}
where $u_{0}$ is  the  mean  energy density of noninteracting photons, equal
to $\pi^{2}/15 \beta^{4}$.

The  mean  energy density $u = U/V$ can now be calculated from (8); the result
is
\begin{equation}
u =\left[1 +{154 \pi^{2} \alpha^{2} \over 2025} \left(T/T_{e}\right)^{4}
\right] u_{0}, \eqnum{9}
\end{equation}
where $T_{e}=m$ is equal  to  5.9 GK.  As expected, the {\it post-Planckian}
correction in Eq. (9) is very small at terrestrially occurring
temperatures,   except   possibly   those   achievable   in  experiments  on
thermonuclear  research.   The  sign  of  the correction is also noteworthy.
Although  the  interaction is attractive, at a given temperature the
interacting  photon  gas has a higher energy density than the
noninteracting one.  To elucidate this result, let us consider the
relation
\begin{equation}
u = u_{0}  + V^{-1} (1 - T \partial /\partial T) {\bar H}_{\text int},
\eqnum{10}
\end{equation}
which is valid to first order in the correction.  Here ${\bar H}_{\text int}$
is the mean value of the interaction Hamiltonian
$tr({\hat H}_{\text int}{\hat \rho})$ in leading
order.  A straightforward calculation now yields
\begin{equation}
{\bar H}_{\text int}= - {22 \pi^{2} \alpha^{2} V\over 2025}
\left(T/T_{e}\right)^{4} u_{0}.\eqnum{11}
\end{equation}
Equation  (10)  shows  that  if,  as  is  the  case  here,
${\bar H}_{\text int}$  depends on the
temperature  more  strongly than linear, then $u$ will exceed $u_{0}$ in
case the interaction is attractive.  We must therefore conclude that the
negative photon-photon interaction energy is in the present instance
more than compensated by an increase in the average number
of photons in each mode, resulting in a net increase in the internal energy.

To  verify  the  increase  in  photon  occupation  number just predicted, we
proceed  to  calculate  the  ensemble average of the photon number operator,
$tr[{\hat n}({\bf k},{\lambda}){\hat \rho}]$.  To the leading order we find,
after some algebra,
\begin{equation}
{\bar n}(k) = \left\{1 + \left[1 + {\bar n}_{0}(k)\right]{44 \pi^{2}
\alpha^{2}k \over 2025 m}
\left(T/T_{e}\right)^{3}\right\}{\bar n}_{0}(k).\eqnum{12}
\end{equation}
This  formula  gives the leading correction to Planck's radiation law arising
from vacuum polarization.
As  predicted  above,  the  correction is an increase in occupation
number for all modes.  Note that the increase is more pronounced for higher
energy modes, reflecting the increase with energy of the
photon-photon scattering cross section.

Other  quantities  of  interest  are  calculated  in  a similar manner.  For
example, the pressure for the photon gas is found to be
\begin{equation}
p = \left[ 1 + {22 \pi^{2} \alpha^{2} \over 675} \left(T/T_{e}\right)
^{4}\right] p_{0}, \eqnum{13}
\end{equation}
where $p(p_{0})$  is  the  pressure for the interacting (noninteracting) case.
Likewise, the speed of sound, given by $dp/du$ in the present case, is
found with the help of
Eqs. (9) and (13).  The result is
\begin{equation}
v_{s}=\left[ 1 - {88 \pi^{2} \alpha^{2} \over 2025} \left( T/T_{e}
 \right)^{4}\right] {c \over \sqrt{3}},\eqnum{14}
\end{equation}
where c is the speed of light in {\it vacuum}.  The speed of light in the
photon gas, on the other hand, is found by first calculating the interaction
energy $\omega_{\text int}$ of an ``external'' photon of momentum $k$ with
the photon gas using Eq. (4).  The result is
\begin{equation}
\omega_{\text int}=-{44 \pi^{2} \alpha^{2} \over 2025} \left(T/T_{e}
\right)^{4}k.\eqnum{15}
\end{equation}
The speed of light in the photon gas, $v_{\gamma}$, is then found
from the formula
$(1+d\omega_{\text int}/dk)c$;
\begin{equation}
v_{\gamma}=\left[1-{44 \pi^{2} \alpha^{2} \over 2025} \left(T/T_{e}
\right)^{4}\right]c. \eqnum{16}
\end{equation}
This result implies that a photon gas in equilibrium acts as a linear,
isotropic refractive medium for the propagation of electromagnetic waves,
its index of refraction being equal to $c/v_{\gamma}$.

As  already  stated, these post-Planckian corrections are very small, even at
high temperatures.  For example,  at  a  temperature of $10^{-1}T_{e}$,
which is of the order of the highest temperatures  presently achievable in
thermonuclear research, the fractional
correction  to  Planck's formula is of the order of $10^{-9}$ (one must also
remember the
contaminations  arising  from the presence of a plasma and the complications
caused  by  the  short-lived  nature  of  the temperature peak in an actual
experiment).  Furthermore, as discussed earlier, the above results should be
supplemented  by  the  condition that requires a photon to collide much more
frequently with another photon than with the equilibrium population of
pair-produced  electrons or positrons.  To find the inequality that enforces
this condition,  we  must  first calculate the photon-photon collision
frequency.

Using  the  standard  result for the number of collisions in a gas,
we  find  for the collision frequency of a given photon,
\begin{equation}
\nu_{\gamma \gamma}(k_{1}) =V^{-1}{\sum}_{2}{\bar n}_{0}(k_{2})
\vert {\hat{\bf k}}_{1}-{\hat {\bf k}}_{2}\vert \sigma ({\bf k}_{1},
{\bf k}_{2}),\eqnum{17}
\end{equation}
where $\sigma ({\bf k}_{1},{\bf k}_{2})$ is  the  scattering
cross section for a pair of photons of momenta ${\bf k}_{1}$ and
${\bf k}_{2}$.  Transforming this expression to the
center-of-mass  system and using the value of $\sigma$ given earlier, we find
\begin{equation}
\nu_{\gamma \gamma}(k) = {2224 \pi^{3} \alpha^{2} \over 455625} r_{e}^{2}
\omega^{3}\left(T/T_{e}\right)^{6}\eqnum{18}
\end{equation}
for  the  collision frequency of a photon of momentum $k$ with the other
photons.

The mean collision  frequency  can  now  be  found  by averaging
$\nu_{\gamma \gamma}(k)$.  The result, stated as a mean free path,
is given by
\begin{equation}
\lambda _{\gamma \gamma} [m] = 1.8 \times 10^{-5} \left( {T[K] \over 5.9
\times 10^{9}} \right) ^{-9}. \eqnum{19}
\end{equation}
This equation gives the mean free path (expressed in meters) in terms of the
temperature  (expressed  in  Kelvins).   Now  the  condition that ensures a
pure photon gas requires that $\lambda _{\gamma \gamma}\ll
\lambda _{\gamma e}$, where $\lambda _{\gamma e}$ is the (partial) mean free
path for the collisions of the photons with pair-produced electrons or
positrons.

A calculation involving the equilibrium concentration of electron-positron
pairs  and  the  Thomson  cross  section  shows that for $T / T_{e}
\ll 1$, the ratio $\lambda _{\gamma \gamma}/ \lambda _{\gamma e}$
is given by $A(T_{e}/T)^{15/2} \exp(-T_{e}/T)$, where $A\simeq5 \times 10^{3}$.
  For $T = 1.5 \times 10^{8}$ K, which is the upper limit to the range of
temperatures considered in this paper, this  ratio  is  less  than
$4\%$, so that the purity condition is well satisfied for the designated range
of temperatures.  Thus for $T \leq 1.5 \times 10^{8}$ K, we have in
Eqs. (8)-(16) the desired post-Planckian corrections to the thermodynamics of
an assembly of photons in equilibrium.

\section{IMPLICATIONS FOR THE COSMIC PHOTON GAS}

At this point the question of the meaning of $V$ in the above derivations must
be  considered  more  carefully  since  the naive model of an enclosure with
impenetrable walls is clearly not tenable at the range of temperatures where
post-Planckian corrections  become  significant.   Indeed  it  is necessary to
examine  the  role of whatever mechanism serves to confine the photon gas to
ensure  that  it does not appreciably affect the thermodynamics, just as was
done  with  pair-produced  electrons and positrons.  As an example, consider
the cosmic photon gas, where one might impose the requirement that the scale
of  inhomogeneities  in  the  cosmic  gravitational  field, which is of the
order of the cosmic scale
parameter  $R$,  be  much  larger  than  the  photon  mean  free  path.   This
requirement basically translates into the inequality $\lambda_{\gamma \gamma }
\ll V^{1/3}$, where $V \sim R^{3}$, a condition that can be interpreted to mean
that photon-photon collisions must be far more frequent than photon-graviton
collisions [12].
Using Eq. (19), we can rewrite the last condition in terms of the
temperature and arrive at the requirements
\begin{equation}
V^{1/3}[m] \gg 4.0 \times 10^{9} \left[ {T[K] \over 1.5 \times 10^{8}}\right]
^{-9}, \ T \le 1.5 \times 10^{8} \ {\text K}.\eqnum{20}
\end{equation}
Observe  that  these  restrictions  on temperature and volume also guarantee
that  on scales larger than $\lambda_{\gamma \gamma}$ the photon gas is in
the {\it hydrodynamic} regime
(in  contrast  to  the {\it independent-particle} regime prevalent under
ordinary conditions), and essentially behaves
like an elastic continuum.  For example, such a gas can support or propagate
density  perturbations of spatial dimensions $d$, provided $V^{1/3} \gg d
\gg \lambda_{\gamma \gamma}$.  It
is  clear,  however,  that  the  necessary conditions for such phenomena are
extreme, suggesting an examination of the early stages of the universe.

According  to  the standard model [13], about one hour after the big bang the
contents  of the universe had cooled to a temperature of about $1.5 \times
10^{8}$ K, and
the spatio-temporal scales of the universe, which for the present discussion
may  be taken to be the Hubble distance and time, were of the order of
$10^{12}$ m and $10^{3}$ s  respectively.   The  requirements  expressed
in Eq. (20)  were thus
satisfied.   Disregarding  for  a  moment  the  rest  of the contents of the
universe   at   this   time,   we  can  see  that  the  photon  gas  behaves
hydrodynamically  on scales of the order of $d$, provided $10^{9}\ {\text m}
 \ll  d \ll 10^{12}$ m.  For example, density perturbations characterized by
wavelengths and frequencies
of  the  order of $10^{10}$ m and $10^{-2}$ s$^{-1}$, respectively, satisfy
these conditions and could have been supported by the cosmic photon gas.
Needless to say, the free
electrons  present in the cosmic fluid at this time (due to the baryon
asymmetry of the universe) cannot be disregarded since
they  play  a  major  role  by virtue of their frequent collisions with the
photons.  A simple calculation shows that at $T \simeq 1.5 \times 10^{8}$ K,
 the ratio $\nu_{\gamma e}/\nu_{\gamma \gamma}$ is  about $10^{4}$, and it
rapidly increases according to $(T/T_{e})^{-6}$ as the universe
cools.    Indeed   by   the   time  the  free  electrons  disappear  due  to
recombination,  the photon gas is essentially noninteracting, as can be seen
by  considering  Eq. (19).  Although at the time of $T \sim 10^{8}$ K
the cosmic photon
gas  was  capable  of  independently  supporting  density  perturbations and
possibly  playing  a  distinct role in structure formation, its dynamics was
largely  dominated  by  the  free electrons ,  thereby  obviating  any
interesting
effects associated with the photons.

It  is  amusing  in  this  connection  to  recall  that  the  phenomenon  of
{\it sonoluminescence},  or  the  creation  of light from sound, was
observed some sixty  years  ago.   The  results  of  this  paper suggest
that the opposite
phenomenon,  which  may  be  named  the {\it luminosonic} effect,  may also
have occurred,  albeit  early  in  the  life  of  the universe, and with
x-rays as ``light'' and infrasonic waves as ``sound.''

\acknowledgments

This work was supported in part by a research award from the California
State University, Sacramento.


\begin{references}
\bibitem[\dag]{email}E-mail address: {\it hpartovi@csus.edu}
\bibitem{1} G. F. Smoot {\it et al.}, Ap. J. {\bf 396}, L13 (1992).
\bibitem{2} O. Halpern, Phys. Rev. {\bf44}, 855 (1933).
\bibitem{3} M. Delbr\"uck, Z. Physik {\bf84}, 144 (1933).
\bibitem{4} R. Wilson, Phys. Rev. {\bf90}, 720 (1953).
\bibitem{5} The interaction of two virtual photons, on the other hand,
is a common
occurrence in present day high-energy collisions; see Ch. Berger and W.
Wagner, Phys. Rep. {\bf146}, 1 (1987).
\bibitem{6} R. Karplus and M. Neumann, Phys. Rev. {\bf83}, 776 (1951).
\bibitem{7} S. Adler, J. Bahcall, C. Callan, and M. Rosenbluth, Phys. Rev.
Lett. {\bf25}, 1061 (1970).  See also
W. Greiner, B. M\"uller and J. Rafelski, {\it Quantum Electrodynamics
of Strong Fields} (Springer-Verlag, Berlin, 1985), p. 291.
\bibitem{8} H. Partovi, ``Detecting Photon-Photon Interaction by Colliding
Laser Beam Interferometry,'' submitted for publication (1993).
\bibitem{9} A. Akhiezer and V. Berestetskii, {\it Quantum Electrodynamics}
(Interscience Publishers, New York,
1965), p. 768.
\bibitem{10} W. Heisenberg and H. Euler, Z. Physik {\bf98}, 714 (1936).
See also V. Weisskopf, Kgl. Danske Videnskab. Selskab, Mat.-fys.
Medd. {\bf 14}, No. 6 (1936), and J. Schwinger, Phys. Rev. {\bf 82},
664 (1951).
\bibitem{11} This is a simple result, although the literature on this point
is often obscure because the field velocity
${\hat{\bf E}}$ is not clearly distinguished from the field momentum
${\hat{\bf \Pi}}$; cf. Ref. [9], p. 780, and V. Weisskopf, Ref. [10], p. 10.
\bibitem{12} Generally,  the  condition
$\lambda_{\gamma \gamma} \ll  V^{1/3}$    is the requirement
that the
photons interact much less with the confining field than with one
another.
\bibitem{13} S.  Weinberg, {\it Gravitation and Cosmology}
(Wiley, New York, 1972); E. Kolb
and M. Turner, {\it The Early Universe} (Addison-Wesley, Reading, MA, 1990).
Note that  the  estimates  given  in  the  text
involve rather large uncertainties characteristic of cosmological data.
\end{references}
\end{document}